\documentclass[a4paper,twocolumn,11pt,accepted=2021-11-29]{quantumarticle}
\pdfoutput=1
\usepackage[utf8]{inputenc}
\usepackage[english]{babel}
\usepackage[T1]{fontenc}
\usepackage{amsmath}
\usepackage{hyperref}
\usepackage{tikz}
\usepackage{lipsum}
\usepackage{bm}
\usepackage{braket}

\begin{document}

\title{Tensor Monopoles in Superconducting Systems}

\author{Hannes Weisbrich}
\affiliation{Fachbereich Physik, Universit{\"a}t Konstanz, D-78457 Konstanz, Germany}
\author{Markus Bestler}
\affiliation{Fachbereich Physik, Universit{\"a}t Konstanz, D-78457 Konstanz, Germany}
\author{Wolfgang Belzig}
\affiliation{Fachbereich Physik, Universit{\"a}t Konstanz, D-78457 Konstanz, Germany}
\email{wolfgang.belzig@uni-konstanz.de}
\homepage{https://www.belzig.uni-konstanz.de}
\maketitle
\begin{abstract}
 Topology in general but also topological objects such as monopoles are a central concept in physics. They are prime examples for the intriguing physics of gauge theories and topological states of matter. Vector monopoles are already frequently discussed such as the well-established Dirac monopole in three dimensions. Less known are tensor monopoles giving rise to tensor gauge fields. Here we report that tensor monopoles can potentially be realized in superconducting multi-terminal systems using the phase differences between superconductors as synthetic dimensions. In a first proposal we suggest a circuit of superconducting islands featuring charge states to realize a tensor monopole. As a second example we propose a triple dot system coupled to multiple superconductors that also gives rise to such a topological structure. All proposals can be implemented with current experimental means and the monopole readily be detected by measuring the quantum geometry.
\end{abstract}
\section{Introduction}
The concept of monopoles was initially introduced in the framework of electrodynamics to prove the quantization of the electric charge \cite{dirac1931quantised}. This idea of the electrodynamic monopoles was later translated to condensed matter using the geometrical properties of Bloch states \cite{qi2011topological}. Here, the central object describing the geometry of these quantum systems in parameter space is the Berry connection with its physical observable: the geometric Berry phase \cite{berry1984quantal}. As the Berry connection has the form of a gauge potential in parameter space one can define a corresponding gauge invariant vector field, that is nothing else than the Berry curvature. Similar as in the electrodynamic case, monopoles can be defined in the parameter space of these quantum systems. Monopoles are sources of the gauge invariant vector field, the Berry curvature, and, hence, a quantized charge can be related to a topological invariant. For instance, in the case of the 3D Dirac monopole the corresponding topological invariant is the first Chern number determining the quantized topological charge \cite{TKKN}. This topological invariant classifies also a wide spectrum of topological states of matter as in topological insulators \cite{Hasan:2010ku}, topological superconductors \cite{Sato:2017go}, or in the quantum Hall effect \cite{klitzing1980new,lhatsugai1993chern}. Indeed the topology emerges in these cases from artificial Dirac monopoles in momentum space. This idea of monopoles can be also extended to higher dimension, for instance in 5D the non-Abelian Yang monopole \cite{yang1978generalization} gives rise to a non-Abelian Berry curvature \cite{wilczek1984app} with its topological charge determined by the second Chern number. Interestingly, such a Yang monopole was realized experimentally in ultracold atoms \cite{sugawa2018second}. More generally the charge of monopoles in $2n+1$ dimensions can be described by the $n$-th Chern number \cite{nakahara2003geometry}.
As these monopoles in odd dimensions give rise to vector gauge fields, they are classified as vector monopoles. 

Another class are tensor monopoles \cite{nepomechie1985magnetic,teitelboim1986m,orland1982instantons}  being sources of tensor gauge fields which are for instance crucial in string theory \cite{banks2011symmetries,mavromatos2017magnetic,montero2017chern}.  The elementary case is the Abelian 4D tensor monopole with its topological charge defined as Dixmier-Douady invariant \cite{mathai2017differential,murray1996bundle}. It was shown that similar as the Berry curvature one can define a tensor Berry curvature \cite{palumbo2019tensor,palumbo2018revealing} to realize fictitious tensor monopoles in quantum systems.
As such tensor monopole require an higher dimensional phase space they can only exist in synthetic dimensions. For instance, it was shown that using the coupling parameters of a superconducting qudrit as artificial dimensions the 4D tensor monopole could be experimentally observed \cite{tan2021experimental}. In a similar experiment a single NV center was used to create this topological structure of a tensor monopole \cite{chen2020synthetic}.
Another prominent approach to topology in synthetic dimensions are multiterminal Josephson junctions \cite{riwar2016multi,eriksson2017topological,xie2017topological,meyer2017nontrivial,xie2018weyl,deb2018josephson,klees2020microwave,xie2019topological,klees2021ground,weisbrich2021second} or topological superconducting circuits \cite{fatemi2021weyl,peyruchat2021transconductance,herrig2020minimal}, where the topology is defined in the space of superconducting phases. Due to the scalability, superconducting Josephson circuits are the most promising candidate for realistic implementations of quantum technologies, see for example the reviews \cite{blais2020quantum,martinis2020quantum}. In the context of topology various topological states of matter can emerge in these systems without dimensional limitations, thus they provide a promising tool to explore the physics of more exotic topological states of matter such as the one of tensor monopoles. 

The goal of this work is to show that tensor monopoles can emerge in such superconducting systems with the monopole defined in the space of the superconducting phases. In the first section we recall the concept of the tensor Berry curvature and tensor monopoles in quantum systems following Ref.\,\cite{palumbo2019tensor,palumbo2018revealing}. In the second section, we describe the superconducting circuit using three superconducting islands to realize a synthetic tensor monopole in the lowest charge states of the circuit. In the last part, we study a system with three dots coupled through superconducting leads.
\begin{figure*}
	\centering
	\includegraphics[width=1\textwidth]{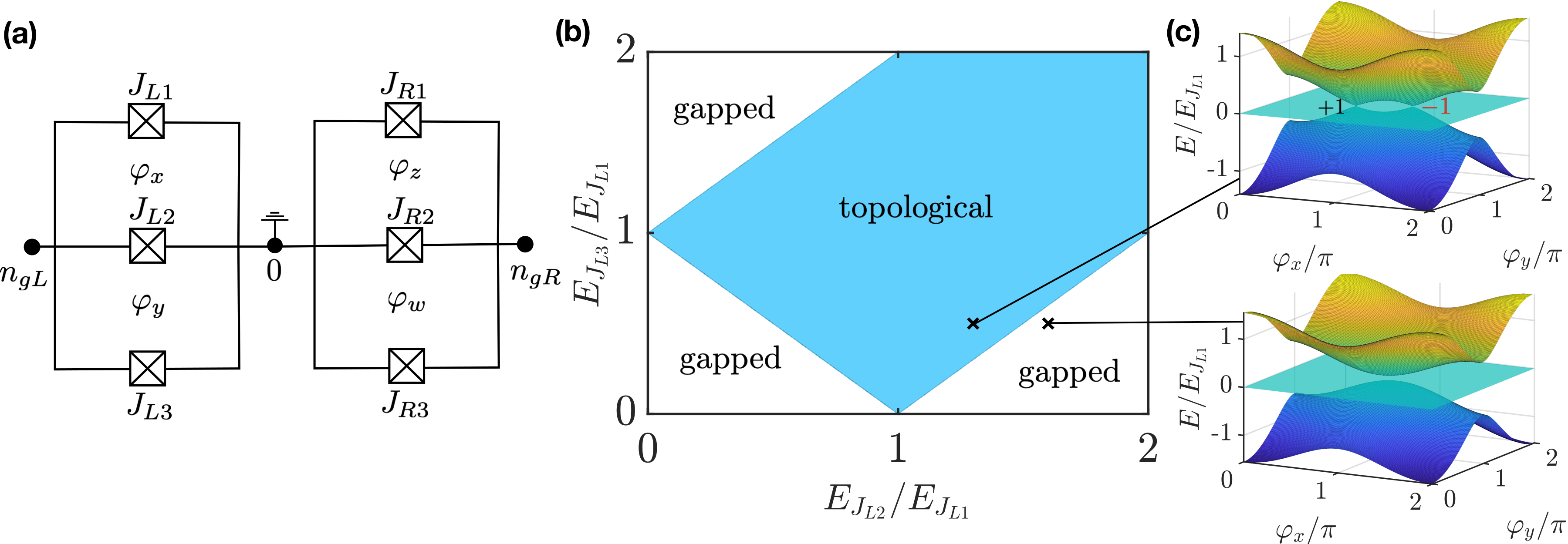}
	\caption{(a) Superconducting circuit with three superconducting islands (black dots) that simulates a tensor monopole using the lowest charge states. The middle superconductor is the reference node, whereas left and right island  are described by their offset charge $n_{gL/R}$. These islands are coupled by several Josephson junctions $J_{L/R\,i}$ (black boxes). The four loops appearing in this circuit are threaded by the magnetic fluxes $\varphi_{x},\varphi_y,\varphi_z,\varphi_w$. (b) Parameter region where tensor monopoles can appear depending on the Josephson energies of the left Josephson junctions. The condition for the right Josephson junctions is equivalent. (c) Energies in the topological region (upper graph at $E_{J_{L3}}=0.5E_{J_{L1}}$ and $E_{J_{L2}}=1.3E_{J_{L1}}$) with the topological charge of the emerging tensor monopoles $Q_T=\pm 1$ and in the gapped region (lower graph at $E_{J_{L3}}=0.5E_{J_{L1}}$ and $E_{J_{L2}}=1.6E_{J_{L1}}$) with $E_{J_{R1}}=E_{J_{R2}}=E_{J_{R3}}$ at $\varphi_z=-1/2\varphi_w=2\pi/3$.}\label{Fig.1}
\end{figure*}
\section{Concept of tensor monopoles}\label{Intro:tensor}
To explain the concept of tensor monopoles we repeat the idea of the better known Dirac monopole in quantum systems. Here the most simple example has the form $H_{\mathrm{3D}}=\bm{q}\bm{\sigma}$ with $\bm{q}=(q_x,q_y,q_z)$ and the Pauli matrices $\bm{\sigma}=(\sigma_x,\sigma_y,\sigma_z)$. This Hamiltonian has two eigenstates with energy $E_{\pm}=\pm|\bm{q}|$. Analogously to the electromagnetic gauge potential one can define the Berry connection $A_{\mu}=i\bra{\psi(\bm{q})}\partial_{q_\mu}\ket{\psi(\bm{q})}$ with $\ket{\psi(\bm{q})}$ being the lower eigenstate. The Berry connection here plays the role of a gauge potential in parameter space $\bm{q}$, whereas the gauge invariant vector field is then the Berry curvature $F_{\mu\nu}=\partial_{q_\mu} A_\nu-\partial_{q_\nu} A_\mu$. To verify this property of gauge invariance one can define the gauge transformation $\ket{\psi}\rightarrow e^{i\gamma}\ket{\psi}$ which transforms the Berry connection as $A_{\mu}=i\bra{\psi}\partial_{q_\mu}\ket{\psi}-\partial_{q_\mu}\gamma$, whereas the additional term cancels out in the Berry curvature. For the above mentioned example the Berry curvature takes the analytical form $F_{\mu\nu}=\epsilon_{\mu\nu\lambda}\frac{q_\lambda}{2|\bm{q}|^3}$. Here the Berry curvature has a singularity at $\bm{q}=0$ when the two eigenstates with $E_{\pm}=\pm |\bm{q}|$ are degenerate which defines the position of the vector monopole in parameter space. Similar as in the electromagnetic case the quantized charge of the monopole can be determined by a closed 2D surface integral of the gauge invariant vector field defined as the first Chern number $C^{(1)}=1/(2\pi)\int_{S^2} F$.\\
This idea of fictitious monopoles in parameter space can be expanded to even dimension with a generalized tensor Berry connection/curvature \cite{palumbo2019tensor,palumbo2018revealing}. Again the idea is best illustrated by a simple example 
\begin{align}
H_{\mathrm{4D}}=\bm{q}\bm{\lambda}=\begin{pmatrix}
0&q_x-iq_y&0\\
q_x+iq_y&0&q_z+iq_w\\
0&q_z-iq_w&0
\end{pmatrix}\label{eq1}
\end{align}
with $\bm{q}=(q_x,q_y,q_z,q_w)$ and $\bm{\lambda}=(\lambda_1,\lambda_2,\lambda_6,\lambda_7^*)$ being four out of eight Gell-Mann-matrices \cite{Gell-mann} with $\text{tr}(\lambda_{i},\lambda_{j})=2\delta_{ij}$ analogously to the lower dimensional Pauli-matrices in the case of the Dirac monopole. This Hamiltonian has in general three eigenstates with $E_0=0$, and $E_{\pm}=\pm|\bm{q}|$. We can see that we have an exceptional Weyl-like point when $|\bm{q}|=0$. The generalized (abelian) tensor Berry connection for the lowest eigenstate $\ket{\psi_{E_-}}=1/\sqrt{2}(v_1,-1,v_2)$ with $v_1=(q_x-iq_y)/|\bm{q}|$ and $v_2=(q_z-iq_w)/|\bm{q}|$ can be defined as \cite{palumbo2019tensor,palumbo2018revealing} 
\begin{align}
	B_{\mu\nu} = \frac{i}{3}\sum_{j,k,l=1}^{3}\epsilon_{jkl} \alpha_j\left(\partial_{q_\mu}\alpha_k\right)\left(\partial_{q_\nu}\alpha_l\right)\label{tensor:con}
\end{align}
with $\alpha_1=-i\log(v_2)$ and $\alpha_2=\alpha_3^*=v_1^*$ and the corresponding tensor Berry curvature
\begin{align}
\mathcal{H}_{\mu\nu\lambda}=\partial_{q_\mu}B_{\nu\lambda}+\partial_{q_\nu}B_{\lambda\mu}+\partial_{q_\lambda}B_{\mu\nu}\, .\label{tensor:curv}
\end{align}  One can verify that under the gauge transformation $\ket{\psi_{E_-}}\rightarrow e^{i\gamma}\ket{\psi_{E_-}}$ the tensor Berry connection transforms as $B_{\mu\nu}\rightarrow B_{\mu\nu}+\Lambda_{\mu\nu}$ with $\Lambda_{\mu\nu}$
being a function of the fields $\alpha_i$ and $\gamma$, whereas the tensor Berry curvature $\mathcal{H}_{\mu\nu\lambda}$ is invariant under a gauge transformation. For the example of Eq. \ref{eq1} the tensor Berry curvature takes the analytical form $\mathcal{H}_{\mu\nu\lambda}=\epsilon_{\mu\nu\lambda\gamma}\frac{q_\gamma}{|\bm{q}|^4}$ indicating the position of the tensor monopole at $|\bm{q}|=0$ when the three states are degenerate. The quantized charge of the tensor monopole can be calculated by the Dixmier-Douady invariant \cite{mathai2017differential,murray1996bundle,palumbo2019tensor} by integrating the tensor Berry curvature over a 3D hyper surface enclosing the tensor monopole
\begin{align}
Q_T=\frac{1}{2\pi^2}\int_{S^3}\mathcal{H}_{\mu\nu\lambda}\,dq^\mu\wedge dq^\nu\wedge dq^\lambda
\end{align}
with $\wedge$ being the wedge product. This results in $Q_{T}=1$ for the discussed example of a single tensor monopole.
\section{Tensor monopoles in a superconducting circuit}

\begin{figure*}
	\centering
	\includegraphics[width=0.8\textwidth]{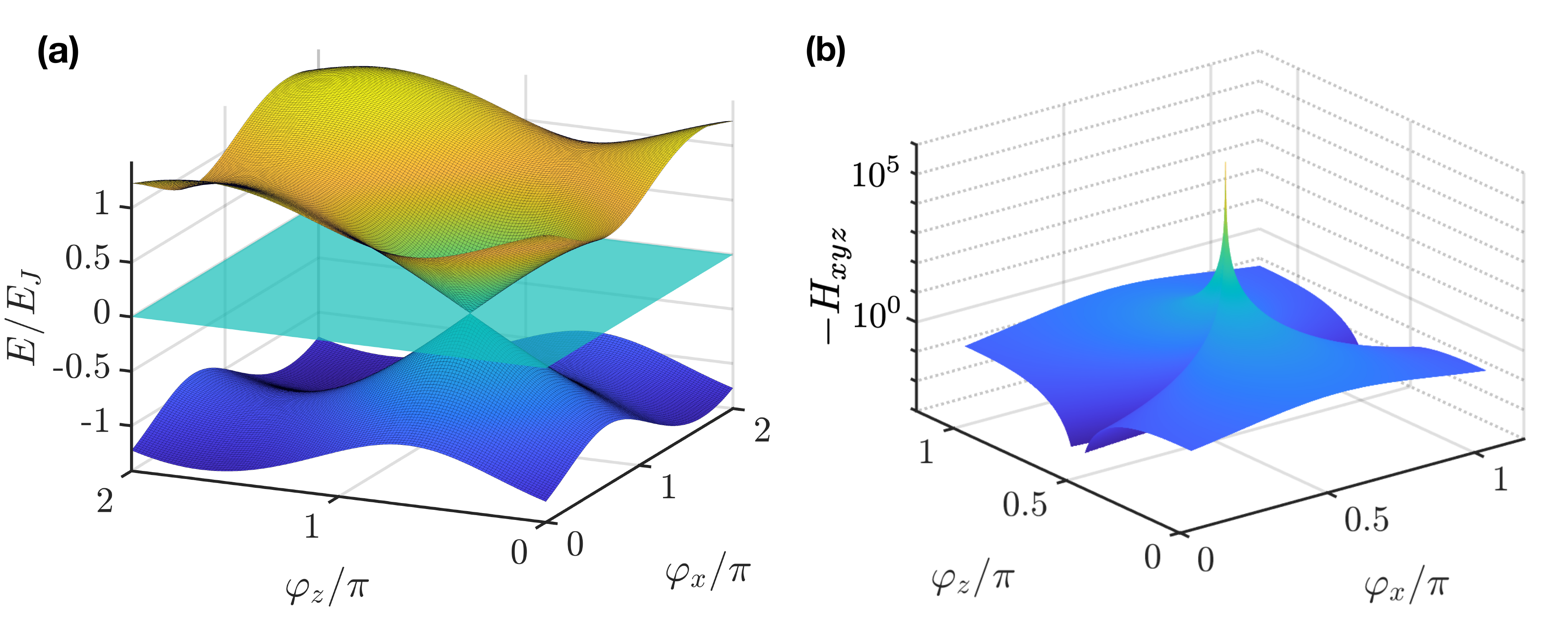}
	\caption{(a) Energy of the three lowest energy levels at $\varphi_y=\varphi_w=-\frac{4\pi}{3}$ with a tensor monopole ($Q_T=-1$) at $\varphi_x=\varphi_z=\frac{2\pi}{3}$ for equal Josephson energies $E_J=E_{J_{L/R\,i}}$. (b) Tensor Berry curvature $\mathcal{H}_{xyz}$ near the the tensor monopole from (a).}\label{tensor_monopole}
\end{figure*}

The system illustrated in Fig.\,\ref{Fig.1}(a) is a superconducting circuit with three superconducting islands (left, middle, and right), whereas the middle island is grounded and connected by six Josephson Junction to the left and right island. We assume that the Josephson Junctions $J_{L/R\,i}$ with $i=1,2,3$ are characterized by their capacitance $C_{L/R\,i}$ and their respective Josephson energy $E_{J_{L/R\,i}}$. Additionally the circuit loops are threaded by the magnetic fluxes $\varphi_x,\varphi_y$ (left) and $\varphi_z,\varphi_w$ (right) which are also used to construct the artificial phase space for the tensor monopoles. Following Ref.\,\cite{vool2017introduction} the Hamiltonian for this system can be in general expressed as
\begin{widetext}
	\begin{align}
		H_{circ}&=\frac{(2e)^2}{2}(\hat{\bm{n}}-\bm{n}_g)^{T}\bm{C}^{-1}(\hat{\bm{n}}-\bm{n}_g)+E_{J_{L\,1}}\cos\left(\hat{\phi}_{L}\right)+E_{J_{L\,2}}\cos\left(\hat{\phi}_{L}-\varphi_x\right)+E_{J_{L\,3}}\cos\left(\hat{\phi}_{L}-\varphi_x-\varphi_y\right)\nonumber\\&+E_{J_{R\,1}}\cos\left(\hat{\phi}_{R}\right)+E_{J_{R\,2}}\cos\left(\hat{\phi}_{R}+\varphi_z\right)+E_{J_{R\,3}}\cos\left(\hat{\phi}_{R}+\varphi_z+\varphi_w\right)
	\end{align}
\end{widetext}
with $\hat{\bm{n}}=(\hat{n}_L,\hat{n}_R)$ the number operator of Cooper pairs on the left and right islands, $\bm{n}_g=(n_{gL},n_{gR})$ are the offset charges of each island controlled by gate voltages, $\hat{\phi}_{L/R}$ are the superconducting phase operators of the left and right superconductors island in respect to the grounded island in the middle, and $E_{J_{L/R\, i}}$ are the Josephson energies of the respective junctions. Note that a specific gauge has been chosen here for the magnetic fluxes. The inverse capacitance matrix $\bm{C}^{-1}$ takes the diagonal form $\bm{C}^{-1}=\text{diag}\left(\frac{1}{C_{L1}+C_{L2}+C_{L3}},\frac{1}{C_{R1}+C_{R2}+C_{R3}}\right)$. For the sake of simplicity we will assume in the following that all junctions have the same capacitance $C\equiv C_{L/R\, i}$. \\
In the charging limit when $E_{C}=(2e)^2/(2C)\gg E_{J}$ the number of charges on the islands is a well defined quantity. Hence, the lowest energy states are the three charge states with zero Cooper pairs on the islands $\ket{0}\equiv\ket{0,0}$ (with $\ket{n_L,n_R}$), one Cooper pair on the left island $\ket{L}\equiv\ket{1,0}$, and one Cooper pair on the right island $\ket{R}\equiv\ket{0,1}$. These charge states can be used to construct the low energy Hamiltonian in this limit
\begin{align}
H_{0}&=\frac{E_C}{3}(1-2n_{gL})\ket{L}\bra{L}+\frac{E_C}{3}(1-2n_{gR})\ket{R}\bra{R}\nonumber\\&-f_L(\varphi_x,\varphi_y)\ket{L}\bra{0}-f_R(\varphi_z,\varphi_w)\ket{0}\bra{R}+h.c.\nonumber
\end{align}
up to a constant and with $f_L(\varphi_x,\varphi_y)=E_{J_{L1}}+E_{J_{L2}}e^{i\varphi_x}+E_{J_{L3}}e^{i(\varphi_x+\varphi_y)}$ and $f_R(\varphi_z,\varphi_w)=E_{J_{R1}}+E_{J_{R2}}e^{i\varphi_z}+E_{J_{R3}}e^{i(\varphi_z+\varphi_w)}$. See Appendix \ref{circuit} for a more detailed derivation of the low energy Hamiltonian.
The energy of the lowest state in general is dominated by the diagonal charging energy, however at the charge degenerate point when $n_{gL}=n_{gR}=1/2$ all three states have the same charging energy, thus the off-diagonal Josephson part dominates the energetic structure. In this case, the Hamiltonian takes a similar form as the tensor monopole in Sec.~\ref{Intro:tensor} and can be expressed with the Gell-Mann matrices $\lambda_i$ in the space $(\ket{L},\ket{0},\ket{R})$ as
\begin{align}
	H_{0}=&\Re(f_L(\varphi_{x},\varphi_y))\lambda_1-\Im(f_L(\varphi_{x},\varphi_y))\lambda_2\nonumber\\&+\Re(f_R(\varphi_{z},\varphi_w))\lambda_6+\Im(f_R(\varphi_{z},\varphi_w))\lambda_7^*\, .
\end{align} 
The energies are $E_{0}=0$, and $E_{\pm}=\pm\sqrt{|f_L(\varphi_x,\varphi_y)|^2+|f_R(\varphi_z,\varphi_w)|^2}$.  Thus it allows for triply degenerate points in the phase space $(\varphi_x,\varphi_y,\varphi_z,\varphi_w)$ similar to the example in Sec.\, \ref{Intro:tensor}. The degenerate points can be determined by the condition $|f_{L}(\varphi_x,\varphi_y)|=|f_{R}(\varphi_z,\varphi_w)|=0$ which depends on the Josephson energies of the individual junctions. In general the condition for $E_{J_{L\, i}}$ and $(\varphi_x,\varphi_y)$ are independent of the condition for  $E_{J_{R\, i}}$ and $(\varphi_z,\varphi_w)$ as the two functions $f_L$ and $f_R$ are independent of each other. As a result of the periodicity of the functions $f_{L/R}$ four single degenerate points can emerge in the phase space if $E_{J_{L\, 2}}+E_{J_{L\, 3}}>E_{J_{L\, 1}}>|E_{J_{L\, 2}}-E_{J_{L\, 3}}|$ and  $E_{J_{R\, 2}}+E_{J_{R\, 3}}>E_{J_{R\, 1}}>|E_{J_{R\, 2}}-E_{J_{R\, 3}}|$, as illustrated in Fig.\,\ref{Fig.1}. If one of these conditions is not fulfilled a gapped phase appears and no tensor monopole can emerge in the system. However, as shown in Fig.\,\ref{Fig.1}, a wide parameter region exists for which these degenerate points occur. 

Using the definition of the tensor Berry connection $B_{\mu\nu}=\frac{i}{3}\sum_{jkl}\epsilon_{jkl}\alpha_j\left(\partial_{\varphi_\mu}\alpha_k\right)\left(\partial_{\varphi_\nu}\alpha_l\right)$ and the tensor Berry curvature
$\mathcal{H}_{\mu\nu\lambda}=\partial_{\varphi_\mu}B_{\nu\lambda}+\partial_{\varphi_\nu}B_{\lambda\mu}+\partial_{\varphi_\lambda}B_{\mu\nu}$ one can verify that in the topological region the degenerate points are indeed tensor monopoles with charges $Q_{T}=\pm 1$. The charge was evaluated numerically by a Monte-Carlo integration of the tensor Berry curvature around a small hypercube around the degenerate points, see Appendix \ref{integral} for the concrete integral. Note that the result is independent of the surface enclosing the degenerate point, e.g. for a hyper spherical surface the integral results in the same topological charge.
In case of equal Josephson energies $E_{J}=E_{J_{L/R\,i}}$, the position of these tensor monopoles are independent of the Josephson energies $E_J$ and have the simple form
$\varphi_{x0}=s_1\frac{2\pi}{3}$,  $\varphi_{y0}=-s_1\frac{4\pi}{3}$, $\varphi_{z0}=s_2\frac{2\pi}{3}$, $\varphi_{w0}=-s_2\frac{4\pi}{3}$
with $s_1=\pm1$ and $s_2=\pm1$. The topological charge of these four tensor monopoles are $Q_{T}=-s_{1}s_{2}$. Another characteristic behavior is the diverging tensor curvature at the monopoles position which can be seen in Fig.~\ref{tensor_monopole} for the example of $\mathcal{H}_{xyz}$, with similar results for the other components of $\mathcal{H}_{\mu\nu\lambda}$. Here, the diverging peak coincides with the position where the three states are degenerate confirming that a tensor monopole is indeed observed.

\section{Tensor monopoles in a triple dot system}

In the following we will discuss another system using three dots coupled to superconducting leads as illustrated in Fig.\,\ref{Fig:4}(a). The Hamiltonian of the dots is $H_{\mathrm{d}}=\sum_{\alpha=L,M,R}\sum_\sigma\left(\epsilon_{\alpha}d_{\alpha\sigma}^\dagger d_{\alpha\sigma}+U n_{\alpha\downarrow}n_{\alpha\uparrow}\right)$ with the fermionic operators $d_{\alpha}$ of the dots (left dot $\alpha=L$, middle dot $\alpha=M$ and right dot $\alpha=R$), $\epsilon_{\alpha}$ the energy of the dot levels, and $U$ the Coulomb energy taken to be same for all dots. The superconducting leads with the phases $\phi_j$ are described by a BCS Hamiltonian $H_{\mathrm{s},j}=\sum_{\bm{k}\sigma}\epsilon_{\bm{k}}c_{j\bm{k}\sigma}^\dagger c_{j\bm{k}\sigma}+\sum_{\bm{k}}\left(\Delta e^{i\phi_j}c_{j\bm{k}\uparrow}^\dagger c_{j-\bm{k}\downarrow}^\dagger+h.c.\right)$, where $\Delta$ is the superconducting gap (assumed to be the same for all leads), $\epsilon_k$ the normal state dispersion, and $c_{j\bm{k}\sigma}$ is the creation operator of an electron with momentum $\bm{k}$ and spin $\sigma$ in the respective lead with phase $\phi_j$. Additionally a normal coupling between the dots $H_{\mathrm{c,d}}=\sum_\sigma v_L d_{L\sigma}^\dagger d_{M\sigma}+v_R d_{M\sigma}^\dagger d_{R\sigma}+h.c.$, a coupling between neighboring dots and leads $H_{\mathrm{c},\mathrm{d}_\alpha-\mathrm{s}_j}=\sum_{\sigma k} v_j \left(d_{\alpha\sigma}^\dagger c_{jk\sigma}+h.c.\right)$, and  a coupling between neighboring superconducting leads $H_{\mathrm{c},\mathrm{s}_i-\mathrm{s}_j}=\sum_{\sigma k} w_{ij} \left(c_{ik\sigma}^\dagger c_{jk\sigma}+h.c.\right)$ is assumed. 
 The total Hamiltonian can then be summarized to \begin{align}
H_{\mathrm{tot}}=&H_{\mathrm{d}}+H_{\mathrm{c,d}}+\sum_{j}H_{\mathrm{s},j}+\sum_{\braket{\alpha,j}}H_{\mathrm{c},\mathrm{d}_\alpha-\mathrm{s}_j}\nonumber\\&+\sum_{\braket{i,j}} H_{\mathrm{c},\mathrm{s}_i-\mathrm{s}_j}\, ,
 \end{align}
 with $\sum_{\braket{\alpha,j}}$ denoting the sum over the three dots ($\alpha=L,M,R$) and their respective nearest leads ($j$) and $\sum_{\braket{i,j}}$ the sum over nearest neighbors of superconducting leads as illustrated in Fig.~\ref{Fig:4}.

\begin{figure}
 	\centering
 	\includegraphics[width=0.5\textwidth]{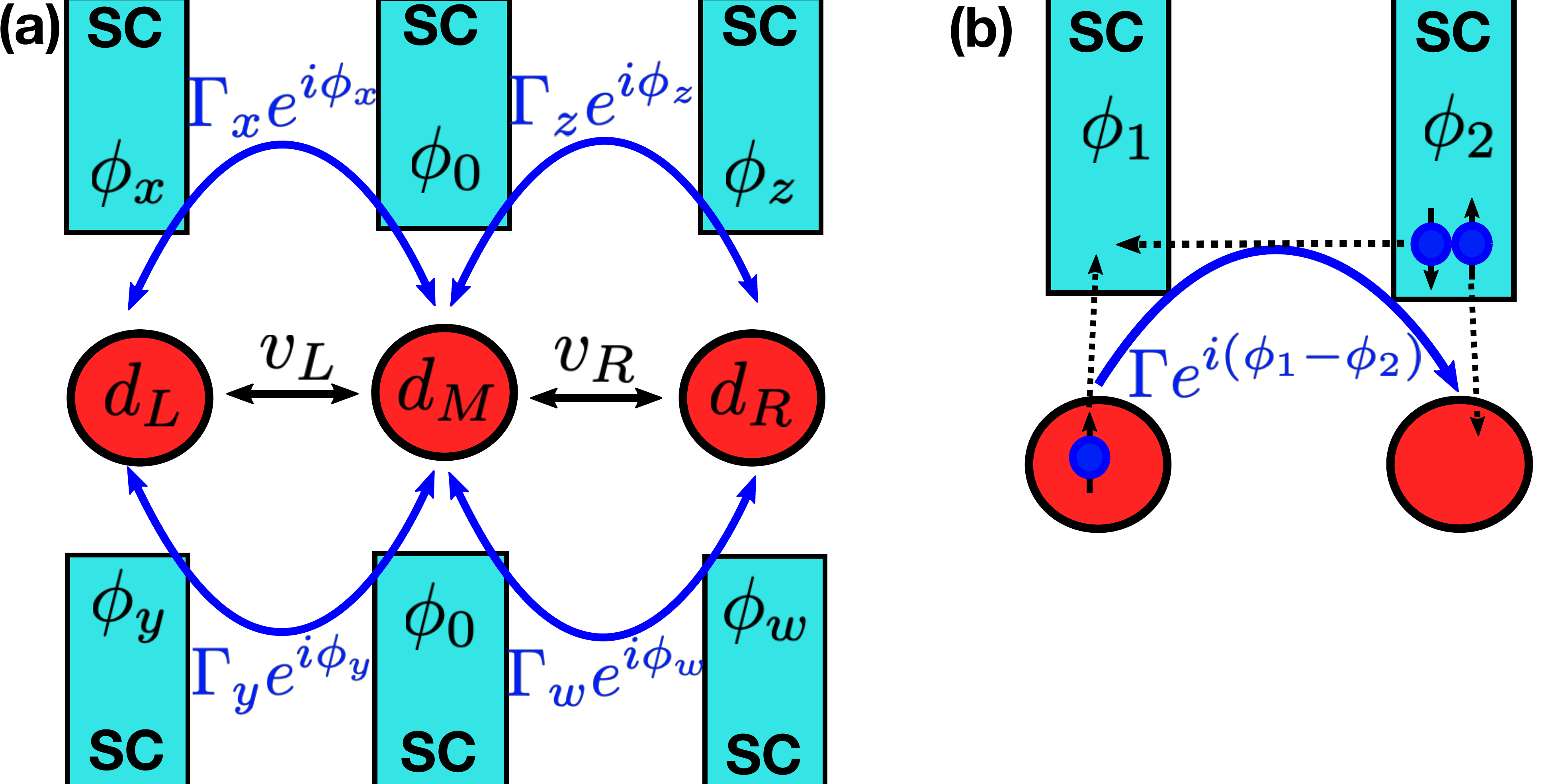}
 	\caption{(a) Three dots coupled via superconducting leads with the phases $\phi_x,\phi_y,\phi_z,\phi_w$ and $\phi_0=0$ as reference phase. Black arrows indicate a normal coupling between dots, whereas blue arrows indicate the indirect coupling via the superconducting leads. (b) This indirect coupling process via the leads is a result of two crossed Andreev reflections where in one lead a new Cooper pair is combined and in the other lead a Cooper pair is split which results in an effective electron transport between two dots.}\label{Fig:3}
\end{figure}

\begin{figure*}
	\centering
	\includegraphics[width=0.99\textwidth]{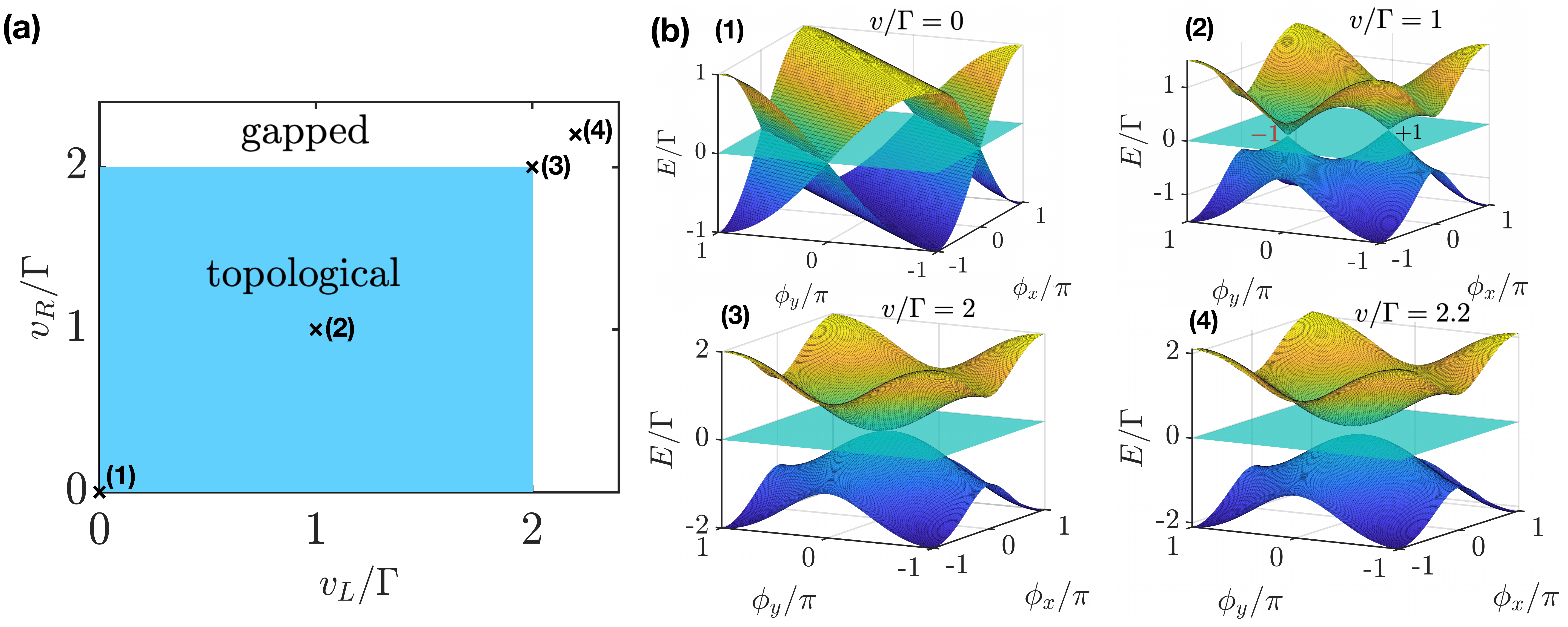}
	\caption{(a) Parameter region for $v_L/\Gamma$ and $v_R/\Gamma$ with four tensor monopoles for equal non-local couplings $\Gamma\equiv \Gamma_i$ in the triple dot system. (b) Energy in dependence of $\phi_x$ and $\phi_y$ for $-\phi_z=\phi_w=\arccos(v/2\Gamma)$, $\epsilon=0$, and for different ratios $v/\Gamma$ ($v=v_L=v_R$) marked in (a) as black crosses. The topological charge of appearing monopoles are indicated by $Q_T=\pm1$.}\label{Fig:4}
\end{figure*}

As we are only interested in the low energy states within the gap, the limit of $\Delta\rightarrow \infty $ is assumed. For a weak coupling between the superconducting leads $w_{ij}\ll v_j$ one can derive an effective Hamiltonian for the dot states by integrating out the superconductors. The details of this calculation are given in the Appendix \ref{green}. In general the coupling between leads and dots can lead to Cooper pair injections on the dots which is however excluded by a large local Coulomb interaction $U$ suppressing double occupation on the dots. An additional effect of the coupling between superconductors and dots is schematically illustrated in Fig.\ref{Fig:3}(b), here the coupling enables a non-local tunneling of electrons via two superconductors. This non-local transport is possible due to two crossed Andreev reflections where in one lead a new Cooper pair is combined and in the other lead a Cooper pair is split which results in an effective transport between two dots. The crucial point of this nonlocal transport is the dependence on the phase difference of the two SC, which allows to construct an effective Hamiltonian for the low energy state with artificial dimensions being the phase differences of the leads.

As a result of these transport processes enabled by the coupling between the dots and the superconductors the low-energy Hamiltonian is conserving electron number and the spin. Hence, we obtain the effective single-particle Hamiltonian
\begin{align}
H_{\mathrm{eff}}&=\sum_{\alpha=L,M,R}\epsilon_\alpha\ket{\alpha_\sigma}\bra{\alpha_\sigma}\nonumber\\&+f_{LM}(\phi_x,\phi_y)\ket{M_\sigma}\bra{L_\sigma}+h.c.\nonumber\\&+f_{MR}(\phi_z,\phi_w)\ket{R_\sigma}\bra{M_\sigma}+h.c\, .\label{eq:eff}
\end{align} 
where we denoted the states $\ket{L_\sigma}\equiv d_{L\sigma}^\dagger\ket{0}$ (with $\ket{0}$ the state without electrons on the dots), $\ket{M_\sigma}\equiv d_{M\sigma}^\dagger\ket{0}$, $\ket{R_\sigma}\equiv d_{R\sigma}^\dagger\ket{0}$, and the total couplings $f_{LM}=v_L-\Gamma_xe^{-i\phi_x}+\Gamma_ye^{-i\phi_y}$, $f_{MR}=v_R-\Gamma_ze^{i\phi_z}+\Gamma_w e^{i\phi_w}$. Note that gauge invariance allows us to set one phase to zero ($\phi_0=0$) as a reference. In addition $\Gamma_j=2\pi N_0 v_jv_0w_{j0}$ are the non-local tunneling amplitudes with $N_0$ the normal density of states at the Fermi energy of the superconductors. For equal energies of the dots $\epsilon_\alpha=\epsilon$ the energy of the lowest states read $E_0=\epsilon$ and $E_{\pm}=\epsilon\pm\sqrt{|f_{LM}|^2+|f_{MR}|^2}$. Hence in general for $\Gamma_x+\Gamma_y>v_L>\big|\Gamma_x-\Gamma_y\big|$ and $\Gamma_z+\Gamma_w>v_R>\big|\Gamma_z-\Gamma_w\big|$ (assuming $\Gamma_i,v_i>0$) four isolated triply degenerate points appear. 

The topological charges of these points can be again determined by an integration of the tensor Berry curvature $\mathcal{H}_{\mu\nu\lambda}$ over a closed surface around the degeneracies which yields the charges $Q_{T}=\pm 1$. Otherwise, if one of the condition is not fulfilled, a gapped phase emerges as illustrated in Fig.~\ref{Fig:4}. In the limit of symmetric couplings $v\equiv v_{L/R}$ and $\Gamma\equiv \Gamma_j$ and $0<v<2\Gamma$ the four triply degenerate point can be observed at $\phi_{x0}=s_1\arccos(v/2\Gamma)$, $\phi_{y0}=-s_1\arccos(v/2\Gamma)$, $\phi_{z0}=s_2\arccos(v/2\Gamma)$, and $\phi_{w0}=-s_2\arccos(v/2\Gamma)$ with $s_1=\pm1$, $s_2=\pm1$ and the topological charge $Q_{T}=s_1s_2$. In the case of $v=0$ degenerate lines in the phase space of  $(\phi_x,\phi_y,\phi_z,\phi_w)$ appear at $\phi_x=\phi_y+\pi$ and $\phi_z=\phi_w+\pi$ which cannot be classified as tensor monopoles as there are no isolated degenerate points in this case. For $2\Gamma<v$ a gapped phase appears and, thus, no monopoles with degeneracies can emerge, as depicted in Fig.\ref{Fig:4}(b). We note, that both a direct coupling between the dots and comparable non-local coupling via the superconductors are necessary to have tensor monopoles in the triple dot chain.

In the regime $0<\frac{v}{\Gamma}<2$ the Hamiltonian can be linearized close to the degenerate points $(\phi_{x0},\phi_{y0},\phi_{z0},\phi_{w0})$ to prove a similar form as in the example of Sec. \ref{Intro:tensor}.  The linearized couplings near these degeneracies are
\begin{align}
 f_{LM}\approx &s_1\Gamma\sqrt{1-\left(\frac{v}{2\Gamma}\right)^2}\left(\delta \phi_x-\delta \phi_y\right)\nonumber\\&-is_1\frac{v}{2}\left(\delta \phi_x+\delta \phi_y\right)\\
 f_{MR}\approx& s_2\Gamma\sqrt{1-\left(\frac{v}{2\Gamma}\right)^2}\left(\delta \phi_z-\delta \phi_w\right)\nonumber\\&+is_2\frac{v}{2}\left(\delta \phi_z+\delta \phi_w\right)
\end{align} 
with $\delta \phi_j=\phi_j-\phi_{j0}$ the detuning of the phases from the position of the tensor monopoles.
Thus it defines the same Hamiltonian of Eq.\ref{eq1} with $q_x\equiv s_1\Gamma\sqrt{1-\left(\frac{v}{2\Gamma}\right)^2}\left(\delta \phi_x-\delta \phi_y\right)$, $q_y\equiv s_1\frac{v}{2}\left(\delta \phi_x+\delta \phi_y\right)$, $q_z\equiv s_2\Gamma\sqrt{1-\left(\frac{v}{2\Gamma}\right)^2}\left(\delta \phi_z-\delta \phi_w\right)$, and $q_w\equiv s_2\frac{v}{2}\left(\delta \phi_z+\delta \phi_w\right)$ ultimately with the same analytical form of the tensor Berry curvature $\mathcal{H}_{\mu\nu\lambda}=\epsilon_{\mu\nu\lambda\gamma}\frac{q_\gamma}{|\bm{q}|^4}$ in the vicinity of the tensor monopoles.

\section{Discussion}

In this work we have theoretically studied tensor monopoles in superconducting systems. These synthetic tensor monopoles which are defined in the 4D-space of the superconducting phases can be characterized by a tensor Berry curvature inspired by the better-known Berry curvature. We have shown that there is a wide variety of approaches in superconducting systems to realize this exotic type of topology. Our first proposal uses the Josephson-coupled charge states of superconducting islands to realize the necessary topological structure. The second example realizes the monopoles in the low energy states of three dots coupled to several superconducting terminal. Remarkably both systems allow to engineer nontrivial topological tensor monopoles in a large parameter region implying robustness against certain fluctuations of the systems parameters. In the example of the superconducting circuit however larger charge fluctuations could be detrimental for the degenerate points in parameter space, as it requires having three degenerate charge states. Similar problems might occur in the second proposal in respect to large asymmetric fluctuations in the dots energies.

In a setup using the coupling parameters of a superconducting qudrit to construct the required artificial 4D space these tensor monopoles were already studied experimentally \cite{tan2021experimental}, as well as in \cite{chen2020synthetic} where a single NV center was observed to have the same topological structure. In both cases the tensor Berry curvature was measured indirectly via the quantum geometric tensor which can be used to reconstruct the tensor Berry curvature \cite{palumbo2018revealing}. In the first experiment the quantum geometric tensor was measured by quenching the system \cite{tan2021experimental}. In the second example the quantum geometric tensor was determined by evaluating the Rabi frequencies for small modulations of the coupling parameters \cite{chen2020synthetic}. The same approaches should be also feasible in the proposed systems. A similar manipulation of the phases could be realized by controlling the magnetic fluxes, either by applying a small periodic modulation or by a sudden quench of the flux. Eventually, the response of the system to these perturbations can be used to measure the tensor Berry curvature in the space of the superconducting phases and ultimately to determine the respective topological invariant.\\
To conclude we emphasize that our proposal to realize tensor monopoles in superconducting nanostructures could stimulate further research in the field of tensor monopoles and the respective topology in superconducting systems and Josephson matter systems. It is the first proposal of using superconducting phases as synthetic dimensions to realize tensor monopoles in superconducting systems. In principle this idea of the tensor Berry connection/curvature and the respective tensor monopole could be expanded to higher-order tensors \cite{palumbo2019tensor}. This seems experimentally accessibly in superconducting systems due to the unlimited dimensions available in the space of superconducting phases providing an alternative path to construct exotic topological states of matter.\\

The authors acknowledge funding provided by the Deutsche Forschungsgemeinschaft (DFG, German Research Foundation) Grant No. RA 2810/1 and SFB 1432 – Project-ID 425217212.

\bibliographystyle{plain}

\onecolumn\newpage
\appendix

\section{Derivation of the low energy Hamiltonian of the superconducting circuit}\label{circuit}
The Hamiltonian for the circuit can be derived with the Lagrangian following Ref. \cite{vool2017introduction} by $\mathcal{L}=\mathcal{L}_{c}-U_J$ with the charging term of the Lagrangian $\mathcal{L}_c$ and $U_J$ the Josephson energy of the junctions.
\begin{align}
\mathcal{L}_c=&\frac{\hbar^2}{8e^2}\left(\sum_{i=1}^{3}C_{Li}\dot{\phi}_L^2+\sum_{i=1}^{3}C_{Ri}\dot{\phi}_R^2\right)=\frac{\hbar^2}{8e^2}\dot{\bm{\phi}}\bm{C}\dot{\bm{\phi}}\\
U_j=&-E_{J_{L1}}\cos(\phi_L)-E_{J_{L2}}\cos(\phi_L-\varphi_x)-E_{J_{L3}}\cos(\phi_L-\varphi_x-\varphi_y)\nonumber\\&-E_{J_{R1}}\cos(-\phi_R)-E_{J_{R2}}\cos(-\phi_R-\varphi_z)-E_{J_{R1}}\cos(-\phi_R-\varphi_z-\varphi_w)
\end{align}
with the capacitance matrix $\bm{C}$, and $\bm{\phi}=(\phi_L,\phi_R)$.
 Assuming gate voltages on each island which results in the offset charge $(n_{gL},n_{gR})$  the Hamiltonian takes the form
\begin{align}
H_{circ}&=\frac{(2e)^2}{2}(\hat{\bm{n}}-\bm{n}_g)^{T}\bm{C}^{-1}(\hat{\bm{n}}-\bm{n}_g)+E_{J_{L\,1}}\cos\left(\hat{\phi}_{L}\right)+E_{J_{L\,2}}\cos\left(\hat{\phi}_{L}-\varphi_x\right)+E_{J_{L\,3}}\cos\left(\hat{\phi}_{L}-\varphi_x-\varphi_y\right)\nonumber\\&+E_{J_{R\,1}}\cos\left(\hat{\phi}_{R}\right)+E_{J_{R\,2}}\cos\left(\hat{\phi}_{R}+\varphi_z\right)+E_{J_{R\,3}}\cos\left(\hat{\phi}_{R}+\varphi_z+\varphi_w\right)
\end{align}
as presented in the main part of the manuscript. In the charging dominated regime $E_{c}\gg E_{J}$ the charging states can be used as basis state with the property of the phase operators \cite{vool2017introduction} $e^{i\hat{\phi}_i}\ket{n_i}=\ket{n_i-1}$ such that it introduces a hopping between the charging states $\ket{L}$,$\ket{0}$, and $\ket{R}$ which depends on the magnetic fluxes
\begin{align}
H_{0}&=\frac{E_C}{3}(1-2n_{gL})\ket{L}\bra{L}+\frac{E_C}{3}(1-2n_{gR})\ket{R}\bra{R}\nonumber\\&-\left(E_{J_{L1}}+E_{J_{L2}}e^{i\varphi_x}+E_{J_{L3}}e^{i(\varphi_x+\varphi_y)}\right)\ket{L}\bra{0}+h.c.\nonumber\\&-\left(E_{J_{R1}}+E_{J_{R2}}e^{-i\varphi_z}+E_{J_{R3}}e^{-i(\varphi_z+\varphi_w)}\right)\ket{0}\bra{R}+h.c.\,.
\end{align}
\section{Integral: Hypercube enclosing the degenerate point}\label{integral}
If the degenerate point is at position $(\varphi_{x0},\varphi_{y0},\varphi_{z0},\varphi_{w0})$ the topological charge can be expressed by the integral
\begin{align}
Q_T&=\frac{1}{2\pi^2}\int_{\varphi_{x0}-a}^{\varphi_{x0}+a}\int_{\varphi_{y0}-a}^{\varphi_{y0}+a}\int_{\varphi_{z0}-a}^{\varphi_{z0}+a} \left(H_{xyz}\big|_{\varphi_w=\varphi_{w0}+a}-H_{xyz}\big|_{\varphi_w=\varphi_{w0}-a}\right)d\varphi_xd\varphi_yd\varphi_z\nonumber\\
&-\frac{1}{2\pi^2}\int_{\varphi_{x0}-a}^{\varphi_{x0}+a}\int_{\varphi_{y0}-a}^{\varphi_{y0}+a}\int_{\varphi_{w0}-a}^{\varphi_{w0}+a} \left(H_{xyw}\big|_{\varphi_z=\varphi_{z0}+a}-H_{xyw}\big|_{\varphi_z=\varphi_{z0}-a}\right)d\varphi_xd\varphi_yd\varphi_w\nonumber\\
&+\frac{1}{2\pi^2}\int_{\varphi_{x0}-a}^{\varphi_{x0}+a}\int_{\varphi_{z0}-a}^{\varphi_{z0}+a}\int_{\varphi_{w0}-a}^{\varphi_{w0}+a}\left( H_{xzw}\big|_{\varphi_y=\varphi_{y0}+a}-H_{xzw}\big|_{\varphi_y=\phi_{y0}-a}\right)d\varphi_xd\varphi_zd\varphi_w\nonumber\\
&-\frac{1}{2\pi^2}\int_{\varphi_{y0}-a}^{\varphi_{y0}+a}\int_{\varphi_{z0}-a}^{\varphi_{z0}+a}\int_{\varphi_{w0}-a}^{\varphi_{w0}+a}\left( H_{yzw}\big|_{\varphi_x=\varphi_{x0}+a}-H_{yzw}\big|_{\varphi_x=\varphi_{x0}-a}\right)d\varphi_yd\varphi_zd\varphi_w\nonumber\\
\end{align}
with $2a$ the length of the hypercube in phase space enclosing the degenerate point.
\section{Derivation of the low energy Hamiltonian of the triple dot system}\label{green}
The starting Hamiltonian for the triple dot system with the superconducting leads reads
\begin{align}
H_{\mathrm{tot}}=H_{\mathrm{d}}+H_{\mathrm{c,d}}+\sum_{j}H_{\mathrm{s},j}+\sum_{\braket{\alpha,j}}H_{\mathrm{c},\mathrm{d}_\alpha-\mathrm{s}_j}+\sum_{\braket{i,j}} H_{\mathrm{c},\mathrm{s}_i-\mathrm{s}_j}\, ,
\end{align}
with the Hamiltonian of the dots
\begin{align}
H_{\mathrm{d}}=\sum_{\alpha=L,M,R}\sum_\sigma\left(\epsilon_{\alpha}d_{\alpha\sigma}^\dagger d_{\alpha\sigma}+U_Cn_{\alpha\downarrow}n_{\alpha\uparrow}\right)\, ,
\end{align}
the coupling between the dots
\begin{align}
H_{\mathrm{c,d}}=\sum_\sigma v_L d_{L\sigma}^\dagger d_{M\sigma}+v_R d_{M\sigma}^\dagger d_{R\sigma}+h.c.\, ,
\end{align}
the BCS Hamiltonian of the superconducting leads
\begin{align}
H_{\mathrm{s},j}=\sum_{\bm{k}\sigma}\epsilon_{\bm{k}}c_{j\bm{k}\sigma}^\dagger c_{j\bm{k}\sigma}+\sum_{\bm{k}}\left(\Delta e^{i\phi_j}c_{j\bm{k}\uparrow}^\dagger c_{j-\bm{k}\downarrow}^\dagger+h.c.\right)\, ,
\end{align}
the coupling between neighboring dots and leads
\begin{align}
H_{\mathrm{c},\mathrm{d}_\alpha-\mathrm{s}_j}=\sum_{\sigma k} v_j \left(d_{\alpha\sigma}^\dagger c_{jk\sigma}+h.c.\right)\, ,
\end{align}
and the coupling between neighboring leads
\begin{align}
H_{\mathrm{c},\mathrm{s}_i-\mathrm{s}_j}=\sum_{\sigma k} w_{ij} \left(c_{ik\sigma}^\dagger c_{jk\sigma}+h.c.\right)\, .
\end{align}
Since the couplings do not depend on the quasi-momentum $\bm{k}$ of the electrons in the leads one can already integrate out $\bm{k}$ to obtain an effective Green's function for the leads in Spin-Nambu space
\begin{align}
g_{s,\phi_j}=-\frac{\pi N_0}{\Delta^2-\epsilon^2}\sigma_0\otimes\left(\epsilon\tau_0+\Delta e^{i\phi_j\tau_3}\tau_1\right)\, ,
\end{align}
where $\sigma_j$ are the Pauli matrices in spin space, $\tau_i$ are the Pauli matrices in Nambu space, and $N_0$ is the density of states in the normal state. In the low energy limit $\epsilon \ll \Delta$ the Green's function can be approximated to
\begin{align}
g_{s,\phi_j}\approx-\pi N_0\left(\sigma_0\otimes e^{i\phi_j\tau_3}\tau_1\right)\, .
\end{align}
For small coupling between the leads $w_{ij}\ll v_{j}$ we expand the dressed Green's function of the dots $G_{d}$ with the Dyson equation up to first order in the coupling between the leads
\begin{align}
G_{d}=g_{d}+g_{d}V_{ds}g_{s}\left(V_{sd}+V_{ss}g_sV_{sd}\right)G_{d}\, ,
\end{align}
with $g_{d}=\text{diag}(g_{L},g_{M},g_{R})$ the bare Green's function of the three dots, $g_{s}=\text{diag}(g_{s,\phi_x},g_{s,\phi_y},g_{s,\phi_0},g_{s,\phi_z},g_{s,\phi_w})$ the bare Green's function of the leads,
\begin{align}
V_{ss}=\begin{pmatrix}
0&0&W_{x0}&0&0\\
0&0&W_{y0}&0&0\\
W_{x0}^\dagger&W_{y0}^\dagger&0&W_{z0}&W_{w0}\\
0&0&W_{z0}^\dagger&0&0\\
0&0&W_{w0}^\dagger&0&0
\end{pmatrix}
\end{align}
the interaction between the superconductors with $W_{ij}=w_{ij}\left(\sigma_0\otimes\tau_3\right)$, and $V_{sd}$ the interaction between dots and leads
\begin{align}
V_{ds}=\begin{pmatrix}
V_{x}&V_y&0&0&0\\
0&0&V_0&0&0\\
0&0&0&V_z&V_w\\
\end{pmatrix}=V_{sd}^\dagger\, ,
\end{align} with $V_{j}=v_j\left(\sigma_0\otimes\tau_3\right)$\, .
Finally we obtain an effective Hamiltonian for the dots $H_{\mathrm{eff}}=H_{\mathrm{d}}+H_{\mathrm{c,d}}+\Sigma_{0}+\Sigma_{1}$ with the zeroth-order contribution
\begin{align}
\Sigma_0&=\frac{1}{2}\begin{pmatrix}
\bm{d}_L^\dagger&\bm{d}_{M}^\dagger&\bm{d}_R^\dagger
\end{pmatrix}V_{ds}g_sV_{sd}\begin{pmatrix}
\bm{d}_L\\\bm{d}_{M}\\\bm{d}_R
\end{pmatrix}\nonumber\\&=\left(\pi N_0 \left(v_{x}^2 e^{i\phi_x}+v_{y}^2 e^{i\phi_y}\right)d_{L\uparrow}^\dagger d_{L\downarrow}^\dagger+\pi N_0 \left(v_{z}^2 e^{i\phi_z}+v_{w}^2 e^{i\phi_w}\right)d_{R\uparrow}^\dagger d_{R\downarrow}^\dagger+\pi N_0 v_{o}^2 e^{i\phi_0}d_{M\uparrow}^\dagger d_{M\downarrow}^\dagger \right)+h.c.
\end{align}
with the spinor $\bm{d}_\alpha^\dagger=\left(d_{\alpha\uparrow}^\dagger,d_{\alpha\downarrow},d_{\alpha\downarrow}^\dagger,-d_{\alpha\uparrow}\right)$, and the first order contribution
\begin{align}
\Sigma_1&=\frac{1}{2}\begin{pmatrix}
\bm{d}_L^\dagger&\bm{d}_{M}^\dagger&\bm{d}_R^\dagger
\end{pmatrix}V_{ds}g_sV_{ss}g_sV_{sd}\begin{pmatrix}
\bm{d}_L\\\bm{d}_{M}\\\bm{d}_R
\end{pmatrix}\nonumber\\&=-\sum_{\sigma}\left(\Gamma_{x}e^{i(\phi_x-\phi_0)}+\Gamma_{y}e^{i(\phi_y-\phi_0)}\right)d_{L\sigma}^\dagger d_{M\sigma}+h.c.-\sum_{\sigma}\left(\Gamma_{z}e^{i(\phi_0-\phi_z)}+\Gamma_{w}e^{i(\phi_0-\phi_w)}\right)d_{M\sigma}^\dagger d_{R\sigma}+h.c.\, ,
\end{align}
with $\Gamma_{j}=2\pi N_0^2v_jv_0w_{j0}$.
Assuming a large Coulomb energy $U_{c}\rightarrow \infty$ the Cooper pair injection terms appearing in $\Sigma_0$ are suppressed as doubly occupied states have to pay the large Coulomb energy. Hence the system conserves total electron number and the effective single electron Hamiltonian can be determined to Eq. \ref{eq:eff}.
\end{document}